\documentclass[pre,aps]{revtex4} 
\usepackage{psfig}
\newcommand{\be}{\begin{equation}} 
\newcommand{\ee}{\end{equation}}

\begin{document}

\title{A Quantitative Clustering Approach to
Ultrametricity in Spin Glasses}

\author{Stefano Ciliberti and Enzo Marinari} 
\affiliation
{Dipartimento di Fisica, 
SMC and UdR1 of INFM, INFN,
Universit\`a di Roma {\em La Sapienza},
P.le A. Moro 2, 00185 Roma, Italy}

\begin{abstract}
We discuss the problem of ultrametricity in mean field spin glasses by
means of a hierarchical clustering algorithm.  We complement the
clustering approach with quantitative testing: we discuss both in some
detail.  We show that the elimination of the (in this context
accidental) spin flip symmetry plays a crucial role in the analysis,
since the symmetry hides the real nature of the data.  We are able to
use in the analysis disorder averaged quantities.  We are able to
exhibit a number of features of the low $T$ phase of the mean field
theory, and to claim that the full hierarchical structure can be
observed without ambiguities only on very large lattice volumes, not
currently accessible by numerical simulations.
\end{abstract}

\date{2003, April 10th}

\maketitle

\section*{Happy Birthday}

This paper is to honor Giovanni Jona-Lasinio birthday. We are grateful
to him since he has taught to us, as to so many other people in Rome
and in other places, a lot of physics and much about the way to love
good physics. Thanks, and Happy Birthday!

\section{Introduction\label{S-INTROD}}

The use of clustering methods to qualify the low temperature phase of
spin glass systems has been recently advocated in a group of very
interesting papers \cite{domany}. It is indeed well known that the
broken phase of mean field spin glasses has a high level of
complexity, that translates statically in Parisi spontaneous Replica
Symmetry Breaking (RSB) and dynamically in a series of dramatic
phenomena that go from a severe critical slowing down $\forall\  T<T_c$
to memory effects, aging phenomena and violations of the
fluctuation-dissipation theorem \cite{books}. 

Ultrametricity of states \cite{ultra} is one of the key features of
the mean field Parisi picture: states of the system turn out to be
endowed by an ultrametric distance, and the phase space is organized
hierarchically. Do finite dimensional spin glass systems share this
properties, and can we find a way to check that? This is an important
issue of the persistent debate \cite{review} about the physics of the
low temperature phase of finite dimensional spin glasses.

Detecting ultrametricity on finite volume systems turns out to be very
difficult \cite{camapa,fraric}: the introduction of constrained Monte
Carlo methods \cite{camapa} and the analysis of the dynamical behavior
of the system \cite{fraric} help only marginally. Finite size effects
are very strong, and make the asymptotic potential emergence of a
hierarchical structure difficult to observe.

Here we introduce some new analysis techniques and we study the
Sherrington-Kirkpatrick (SK) mean field model, where we know that for
low $T$ a non-trivial ultrametric structure emerges in the infinite
volume limit. We will find out that this is a difficult task, sharing
all the problems one observes in finite dimensional systems
\cite{domany,camapa}. Our main points can be summarized in four basic
issues:

\begin{enumerate}

\item We find that to be of better use
the approach based on hierarchical clustering has to be 
complemented by the use of testing techniques that have been developed
in the field of numerical taxonomy \cite{jaidub}. We discuss some of
these techniques and we show how they can be applied to our problem.

\item We discuss the role of the $Z_2$ symmetry of the phase space. We
find that removing this symmetry (that in this context is accidental)
is crucial to get sensible results from quantitative tests. We
introduce and discuss the way to remove the symmetry from equilibrium
configurations obtained in zero magnetic field.

\item Thanks to these techniques we are able to clarify how a finite
volume SK system behaves as far as ultrametricity is concerned, by
working out strengths and limitations of the method. We find that on
the (medium-large) lattice sizes that we are able to analyze one can
establish that a structure is emerging, but that one cannot get a
compulsory evidence about this structure being ultrametric. This is
exactly the same kind of phenomenon one observes when studying finite
dimensional systems \cite{domany}.

\item We analyze systematically finite size effects (by studying
systems on different lattice sizes) and the dependence of our results
over $T$. Thanks to the quantitative analysis techniques that we
introduce we are able to use hierarchical clustering techniques to
discuss also quantities that are {\em averaged over the disorder},
opening in this way a large information window.

\end{enumerate}

The low temperature mean field behavior of spin glass systems is
understood in the framework of the Parisi RSB scheme \cite{books}.
The prototype of mean field spin glass models is the SK fully
connected Ising model where coupling constants are \emph{quenched}
random variables:
\begin{equation}
{\mathcal{H}}_J[\sigma]=-\sum_{i,k=1}^N \sigma_i J_{i,k}\sigma_k \ ,
\label{E-H}
\end{equation}
where $\sigma_i=\pm 1$ are spin variables and the $J_{i,k}$ are
distributed according to an even distribution function.  For example
we can use a Gaussian distribution with $\overline{J_{ik}}=0$ (since
we want to avoid ferromagnetic effects) and
$\overline{J^2_{ik}}=\frac1N$ (to ensure that the energy is
extensive). As we have already reminded, the Parisi RSB solution of the
SK model, which is believed to be the correct solution of mean field
theory at low $T$, exhibits an ultrametric organization of the states
\cite{ultra}. This means that in the infinite volume limit for any
triple of equilibrium spin configurations $\alpha,\beta,\gamma$ we
have that:
\begin{displaymath} 
q_{\alpha\beta}\geq \min
\{q_{\alpha\gamma}, q_{\beta\gamma}\} \ ,
\end{displaymath} 
where $q_{\alpha\beta}$ is the overlap among
configurations $\alpha$ and $\beta$, defined as 
\begin{equation}
\label{E-OVERLAP} 
q_{\alpha\beta}\equiv\frac1{N}\sum_{i=1}^{N}
\sigma_i^\alpha \sigma_i^\beta 
\end{equation}
(here configurations $\alpha$ and $\beta$ are independent
configurations at equilibrium under the same Hamiltonian, sharing the
same quenched realization of the random couplings: they are only
coupled by the fact of sharing the same realization of the random
Hamiltonian). The overlap $q_{\alpha\beta}$ is a similarity index, and
the distance is connected to one minus the overlap.

We will analyze in detail the fact that revealing numerically an
ultrametrical emerging structure on finite systems is difficult. The
question is even more relevant since detecting reliable 
signs of an ultrametric structure could be crucial in finite
dimensional systems, where the behavior of the system in the low $T$
phase is not yet understood \cite{review}.

Clustering \cite{jaidub} is a powerful technique for analyzing data
(for interesting applications of statistical mechanical ideas to
clustering see \cite{rogufo,blwido,stibia}).
Since producing a valid hierarchical clustering is equivalent to show
the existence of a true ultrametric structure of the data, this kind of
approach can give crucial evidences. We will discuss here what happens
in the infinite range mean field SK model, where we know that
eventually, in the infinite volume limit, ultrametricity of states
will emerge. We believe this is needed to help in interpreting the
results obtained in the analysis of finite dimensional models
\cite{domany}. We will see that some important hints do indeed
emerge. 

In this note we introduce some new ideas relevant for hierarchical
cluster as applied to the analysis of disordered and complex systems,
and we discuss numerical results obtained from a clustering analysis
of equilibrium spin glass configurations, with a particular emphasis
on the study of the ultrametric nature of these states.  We explain
why a detailed analysis requires an appropriate elimination of the
spin flip symmetry and we investigate the dependence of our results on
the number of degrees of freedom of the system, showing that finite
size effects are actually very large.

The paper is organized as follows. In section \ref{S-CLUSTER} we
introduce the clustering procedure and we explain the motivations for
our precise choice of a given clustering algorithm.  In section
\ref{S-ANALYSIS} we apply this technique to the SK model; we discuss
our findings about ultrametricity, also by comparing them with those
that one obtains by using standard techniques.  Here we will introduce
and use quantitative ways to state the significance of the results
obtained by clustering (mainly in section \ref{SS-QT}).  As we said
before a more detailed analysis requires a previous elimination of the
$Z_2$ symmetry, and this is done in section \ref{SS-REVERSE}: in
section \ref{SS-OTHER} we will also say a few words about using
different clustering schemes. Section \ref{S-SPINS} is dedicated to
the clustering of the spins.  We report our conclusions in the last
section.

\section{The Clustering Algorithm\label{S-CLUSTER}}

Clustering is a special kind of (potentially very powerful)
classification tool. We will give here only the basic 
informations we need for our analysis, and we advise the reader to look
at \cite{jaidub}  for further details.

Let us consider a sample done of $M$ data $x^\mu$, where each data
point $x^\mu\equiv\{ x_1^\mu,\ldots x_N^\mu\}$ is a vector in a
$N$-dimensional space.  We want to study the underlying organization
of the data, i.e. we want to find out whether the data are organized
according to some non-trivial structure.  A problem of this type is
strictly related to pattern recognition analysis and to Bayes decision
theory \cite{dudhar}: it is of very general interest, since it emerges
in many relevant contexts.

The main ingredient for the analysis is the {\em proximity matrix}
$d_{\mu\nu}\equiv d(x^\mu,x^\nu)$.  $d(x^\mu,x^\nu)$ is some measure
of the dissimilarity of data $\mu$ and $\nu$. It is such that
$d_{\mu\mu}=0$ and $d_{\mu\nu}=d_{\nu\mu}\geq 0$.  $d$ does not need
to be a distance (for example the triangular inequality could not
be satisfied) but usually it is one.

By clustering we group the data in sets that can be related among them
in different ways. Here we will use the exclusive (each data belongs
to exactly one cluster), intrinsic (i.e. based only on the proximity
matrix $d$) classification known as {\em hierarchical clustering}.
Hierarchical clustering is a nested sequence of partitions obtained
through a classification technique based on one of many possible
algorithms.  The output of the algorithm can be represented by a
hierarchical tree (a so-called {\em dendogram}). 

A generic (even random) set of data can always be arranged to fit a
tree-like structure: this is indeed what clustering does. After doing
such (potentially arbitrary) clustering we are left with the relevant
question of deciding if the hierarchical structure that has been
reconstructed was somehow intrinsic to the data set: this requires an
analysis \emph{a posteriori}.

So, in hierarchical clustering we start from a set of data, we group
them by some algorithm (that we will specify in the following)
building in this way a hierarchical tree. Comparison of this tree and
the original data can lead to quantitative conclusions about the
presence of a true hierarchical structure in the data.

In the course of a cluster analysis one usually faces two main
problems.

\begin{itemize}

\item The first important step is the definition of the dissimilarity
index $d_{\mu\nu}$ which is not always naturally induced from the
context (data do not necessarily belong to an Euclidean space).

In our case this is an easy problem. Starting from the usual notion of
overlap (\ref{E-OVERLAP}) the distance between two spin configurations
can be for example naturally and easily defined as
\begin{displaymath}
d_{\mu\nu}\equiv\frac{1-q_{\mu\nu}}{2}\ .
\end{displaymath}

\item The second problem is how to update distances among
elements. When we fuse elements $\alpha$ and $\beta$ in element
$\gamma$ (so joining two smaller clusters in a larger one) we have to
define all distances from the new cluster $\gamma$ to all other
clusters of the system $\eta$.  This step is crucial since it can play
a dramatic role in the structure of the iteration, even if in
situation where hierarchical clustering turns out to be {\em natural},
i.e. an intrinsic property of the data set, results have to be
independent from this issue (there exist alternative approaches which
allows to avoid such an explicit choice by means of a priori
hypothesis \cite{blwido,giamar}).

The most part of our results has been obtained by the \emph{Ward
method} (or \emph{minimum variance method}) \cite{ward,jaidub}.  The
method is based on minimizing the square error, and is empirically
known to outperform other hierarchical clustering methods.

When we merge the two clusters that have the smaller distance we 
define the new distance using the following rule:
if $\rho$ and
$\sigma$ merge to form $\rho'$,  
and $n_\alpha$ is the number of elements in the cluster $\alpha$,
then for any other cluster $\tau$:
\begin{equation}
d_{\tau\rho'}=\frac{
(n_\tau+n_\rho)d_{\tau\rho}+
(n_\tau+n_\sigma)d_{\tau\sigma}-
(n_\rho+n_\sigma)d_{\rho\sigma}
}
{n_\tau+n_\rho+n_\sigma}\ .
\label{E-WARD}
\end{equation}
Let ${\mathcal{C}_\alpha}$ stand for one of the clusters of the system 
and consider the quantity
\begin{displaymath}
S=\sum_{\mathcal{C}_\alpha}\tau(\alpha)\ ,
\end{displaymath}
where the sum is over all the clusters defined in the system
and where
\begin{equation}
\tau(\alpha)=\sum_{\;\mu,\nu\in
\mathcal{C}_\alpha} d_{\mu\nu}^2\ .
\label{E-TAU}
\end{equation}
The choice of the Ward algorithm ensures that when merging two
clusters to form a new one $S$ increases of a minimal amount. In other
terms this definition of distance is the one induced from the maximum
likelihood principle.
\end{itemize}
This defines the clustering scheme that we will follow. We will
discuss next how these ideas can be applied to mean field spin glass
models, and how the result can be understood and quantified by testing
the cluster validity.

\section{Cluster Analysis of the SK Mean Field Spin Glass\label{S-ANALYSIS}}

As we have said we have decided to analyze numerically the mean field
SK model.  Since here the infinite volume scenario is under full
control we believe this is a crucial step in understanding what one can
learn from numerical simulations on finite lattices, and to control the
consequences of such results obtained on finite dimensional models
\cite{domany} where, on the contrary, the theoretical scenario is far
from clear.

We have started by generating by an optimized Monte Carlo method a
large number of uncorrelated spin configurations on lattices of
different sizes and for a number of different realizations of the
quenched disorder (on which we eventually average), under the
Hamiltonian (\ref{E-H}), with quenched random couplings assigned under
a Gaussian distribution. We analyze systems with $N$, number of spins,
equal to $128$, $256$ and $512$ ($N=512$ is typical of a medium size
numerical simulation, corresponding for example to a linear size of $8$
in three dimensions). We thermalize our systems at a set of different
values of the temperature typically going from $0.1\ T_c$ (a very low
value, that we can reach only thanks the power of parallel tempering
\cite{PT,PTREV}), and in all cases we analyze $20$ different
realizations of the quenched couplings. For all lattice sizes,
relevant temperatures and disorder realizations we first thermalize
the system.  After doing that we record one spin configuration after
any new set of $1000$ combined full Monte Carlo sweeps and parallel
tempering updates of the system.  The large ``computer time''
separation among different configurations guarantees a very high level
of statistical independence. Residual possible (very small)
correlations would not spoil our analysis but would only make it a bit
less effective. We have recorded $1024$ such independent spin
configuration for each value of the parameters: such configurations
are the basic set of objects that we have clustered.

Parallel tempering \cite{PT,PTREV} has been crucial in allowing to
bring at thermal equilibrium spin configurations at such low
temperature values on acceptable lattice volumes.  The method is based
on simulating in parallel copies of the system at different
temperature $T$ values, allowing the different copies to swap $T$
among them (with a standard Metropolis weight).  This reduces the free
energy barriers, always keeping the different copies at Boltzmann
equilibrium: tempering can be seen as an annealing where the basic
quantity is not energy but free energy.

We have used all standard criteria to check that, when using the
Parallel Tempering optimized Monte Carlo scheme, we have really
reached thermal equilibrium \cite{PTREV}: we have checked that our
sample dependent overlap probability distributions $P_J(q)$ are indeed
well symmetric under $q\longrightarrow -q$, we have checked that all
copies of the system have visited a number of times all available
temperature values, we have checked that the acceptance factor of the
temperature acceptance swap has been of order $0.5$.

In the rest of this note we will work on {\em clustering} these
configurations and on using quantitative testing to extract the
implications of the hierarchical structure that we obtain.

\begin{figure}
\centerline{\psfig{figure=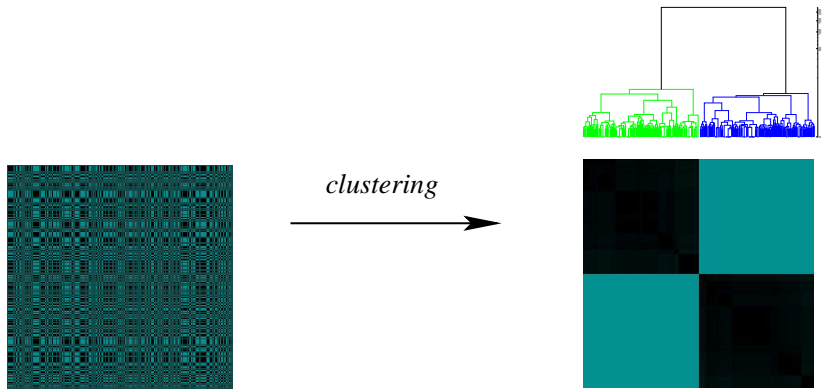,width=0.7\textwidth,angle=90}}
\caption{An example of the clustering procedure as applied to a very
low temperature set of configurations.  In the left part of the figure
we show a proximity matrix $\cal P$ built over $M=512$ configurations
of $N=512$ spins at $T=0.1\ T_c$, ordered at random. 
Darker colors correspond to smaller distances.  On the right
part of the figure we draw the dendogram that results from our
clustering, and the resulting $\cal P$.  
The distance on the dendogram is proportional to $\tau(\alpha)$
defined in equation (\protect\ref{E-TAU}).
The method recovers very well
the structure of two giant clusters related by the $Z_2$ symmetry.}
\label{F-MATRIX-A}
\end{figure}

We first introduce a standard graphical way to get a qualitative
feeling about the set of data. We consider the proximity matrix $\cal
P$, where we have the set of data (in some order to be specified) on
the $x$ and on the $y$ axis, and where we plot with darker colors
points with higher overlap: the diagonal constitutes by definition the
darkest set of the matrix. In figure \ref{F-MATRIX-A} we start by
showing, on the left, the matrix $\cal P$ for a given disorder
realization at $N=512$ and $T=0.1\ T_c$ (a very low value of $T$, the
lowest we have analyzed: here the system is basically in its ground
state) where configurations have been ordered at random.  A clearly
random pattern emerges.

\begin{figure}
\centerline{\psfig{figure=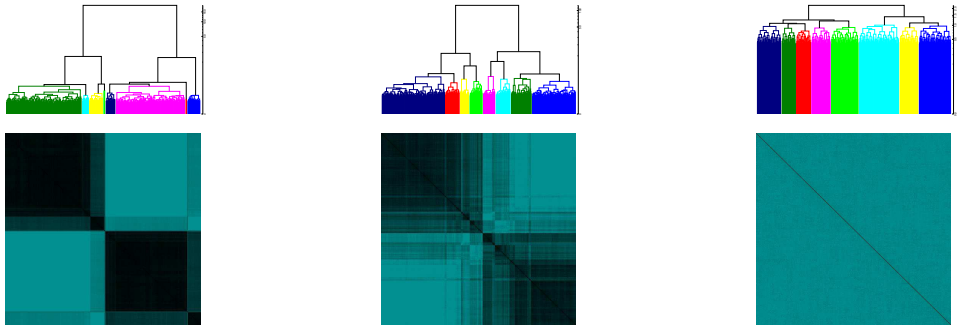,width=0.8\textwidth,angle=90}}
\caption{The dendogram and the related $\cal P$ matrix obtained
from the clustering of $M=256$ configurations at three different
temperature values.  On the left $T=0.3\ T_c$ (where $T$ is very low
and no significant structure but the $Z_2$ degeneracy can be
observed), in the center $T=0.5\ T_c$ (that is the best $T$ region for
observing the non-trivial state structure), and on the right $T=2.0\
T_c$, where there is no structure since we are deep in the high $T$
phase.}
\label{F-3T}
\end{figure}

We apply the Ward algorithm to these configurations in order to obtain a
hierarchical tree (as we have discussed before) \footnote{For
clustering we have used the very flexible set of programs developed
by P. Kleiweg, available from {\tt http://
odur.let.rug.nl/$\tilde{\ }$kleiweg/clustering/clustering.html }}.  The
hierarchical tree that contains the information about the clustering,
the so-called {\em dendogram} \footnote{In a dendogram longer lines are
for farer clusters. In most of our drawings, when we are not
interested in analyzing this specific information, we use an
appropriate power law deformation of the scale to make the graph more
readable and telling.}  is shown in the upper part of the right side
of figure \ref{F-MATRIX-A}.  In the lower part of the right hand side
of figure \ref{F-MATRIX-A} we show the matrix obtained by ordering the
configurations \emph{as from the dendogram} on the $x$ and on the $y$
axis. Now the two reflected states appear very clearly (at such a low
$T$ value there are basically two $\delta$ functions at values $\pm
\overline{q}$, where $\overline{q}$ is close to one). We cannot
observe any further structure, since $T$ is too low (the ideal
temperature value for observing hints of ultrametric effects will turn
out to be, for our lattice sizes, of the order of $0.5\ T_c$).  As we
increase the temperature we observe that well defined structures
emerge (see figure \ref{F-3T}, where we show results for a single
sample, with $N=512$, at $T=0.3\ T_c$, $T=0.5\ T_c$ and $T=2.0\ T_c$):
when we reach the critical temperature $T_c$ and we go deeper in the
warm phase we obtain \emph{a homogeneous matrix}: here spins are
equally likely to be up or down, and as a consequence the overlap
between two configurations is zero on average.

We stress that the information about the $Z_2$ symmetry is a trivial,
well known one, that does not give us further insight: still, it is
interesting that the clustering algorithm is able to reconstruct it.
We will discuss at length the fact that, on the opposite side, the
presence of the symmetry is deeply annoying in that it makes more
difficult to get quantitative information about the structure in one
of the two $Z_2$ sectors, hiding many features of the data, and making
interesting predictions impossible. 

We also use figure \ref{F-3T} to make a further point. The dendograms,
that make possible to visualize the hierarchical structure build from
the clustering, do not give much unambiguous information about the
underlying structure. The picture from $T=0.3\ T_c$ is not so
different, but for some power rescaling of the lengths, from the one
at very high $T$ ($T=2.0\ T_c$) where we do not expect a non trivial
ultrametricity to appear. Clusters at hight $T$ are, indeed, more
balanced, but one can only get some qualitative feelings about it.

In the following we will work on trying to quantify the qualitative
statements about the possible presence of a (maybe hierarchical)
definite structure in the low $T$ phase.

\subsection{Quantitative Testing\label{SS-QT}}

Before discussing our approach toward a quantitative analysis based
on hierarchical clustering techniques and aimed to check whether the
spin configurations (our original data set) are really organized
according to an ultrametric structure, we analyze the system by
applying a more standard statistical mechanical approach.  Following
\cite{domany,camapa} we analyze the probability distribution of the
variable
\begin{displaymath}
k_{\mu\nu\rho}\equiv \frac{d_{\mu\nu}-d_{\mu\rho}} {d_{\nu\rho}} \ ,
\end{displaymath}
where we have ordered the three distances to satisfy the condition
$C\equiv\{d_{\mu\nu}\geq d_{\mu\rho}\geq d_{\nu\rho}\}$.  This implies
that $K\in [0,1]$. In an ultrametric space we would get that
$P(k=K|C)=\delta(k)$.

\begin{figure}
\centerline{\psfig{figure=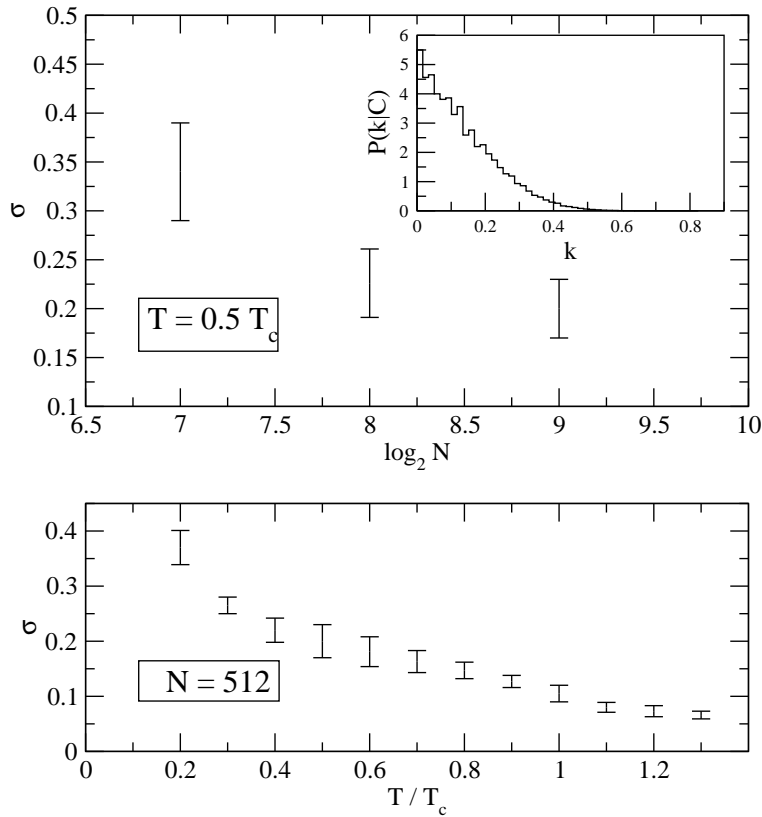,width=0.7\textwidth,angle=270}}
\caption{In the upper part of the figure we plot variance of
the distribution $P(k=K|C)$ versus the base two logarithm of the
number of spins $N$ at fixed temperature $T=0.5\ T_c$.  In the inset
we plot $P(k=K|C)$ as a function of $K$ for a single sample of the
quenched disorder.  In the lower part of the figure we plot $\sigma$
vs. $\frac{T}{T_c}$ for $N=512$.}
\label{F-SIGMA}
\end{figure}

On our finite $N$ lattices we assume the following dependence of $P$
over $K$:
\begin{displaymath}
P(k=K|C) \sim \exp\left\{-\frac{K^2}{2\sigma^2}\right\}\ \theta(K)\ ,
\end{displaymath} 
where $\theta(\cdot)$ is the step function.  We analyze the behavior
of the variance $\sigma^2$ with the size $N$.  We show our results in
figure \ref{F-SIGMA}. In the upper plot we select $T=0.5\ T_c$ and we
plot $\sigma$ as a function of $N$. $\sigma$ decreases with $N$, but
very slowly (as we expected from the results of \cite{camapa}, where
even with a tuned up Monte Carlo procedure one finds that similar
analysis are very difficult): it is not even easy to get a reasonable
fit to a zero limit of $\sigma$ (but the very large statistical error
allows for it). In the inset of the upper part of the plot we show
$P(k=K|C)$ for one single sample. In the bottom plot we show how
$\sigma$ depends on $T$ on our largest lattice size, $N=512$. Nothing
dramatic happens when increasing $T$: again, only some qualitative low
key effect is taking place.

Now we start with analysis of the results of our cluster
reconstruction. We have used our data (spin configurations for a given
lattice size and temperature, together with their mutual distances
obtained from their mutual overlap) to produce a hierarchical tree,
and we want to test if this tree is connected to intrinsic properties
of our data (as we have already clarified an ultrametric tree can
always be superimposed even to random data). We will adapt standard
techniques \cite{jaidub} to judge about the validity of the structure
we have found and about the statement that data are organized
according to an ultrametric structure.

The general procedure testing has a simple structure: given a starting
proximity matrix $\cal P$, we end our clustering procedure with a
particular ordering of elements of $\cal P$, i.e. with a particular
permutation of $|P|$ data. This is what our clustering scheme achieves
(transforming the left part of figure \ref{F-MATRIX-A} in the right
bottom matrix). Now we have the problem of deciding if what we did was
sensible: we can rephrase this question by saying that we have to
choose between the {\em randomness hypothesis} ($H_0$: all
permutations of labels of $M$ are equally likely) and the {\em
alternative hypothesis} ($H_1$: the data have some structure that has
been at least partially reconstructed by the clustering). In order to
check that we:
\begin{enumerate}
\item define a variable $T$ that we expect to be
``small'' under the null hypothesis $H_0$;
\item assign a {\em confidence level} $\alpha$ for $H_1$ and
define a threshold $t_\alpha$ by solving the equation 
$$
P(T\ge t_\alpha | H_0) = 1-\alpha\ ;
$$ 
\item measure from the data
the value of $T$, that we call  $t^*$. If
\begin{enumerate} 
\item $t^* \ge t_\alpha$ $\Rightarrow$ reject $\ H_0$ at level $\alpha$;
\item $t^* < t_\alpha$ $\Rightarrow$ accept $H_0$ at level $\alpha$.
\end{enumerate}
\end{enumerate}
$\alpha$ is a confidence level, i.e. it is connected to
the probability that by accepting $H_1$ as true we are not
making a mistake.

The first tool that we introduce is based on \emph{Hubert's $\Gamma$
Statistics} \cite{jaidub,hubsch}, and it is useful to validate
clustering. This is done by checking the correlation of the data with
a structure we define {\em a priori}.

We consider our measured distance matrix $d_{\mu\nu}$, and we
introduce the matrix $f_{\mu\nu}$ by
\begin{equation}
f_{\mu,\nu}=
\left\{
\begin{array}{cl}
0 & \textrm{if $\mu,\nu\,{{\in}}$ 
same cluster}\\
1 & \textrm{if not}
\end{array}
\right.
\label{E-HUBERT}
\end{equation}
We will study the correlations among  $d_{\mu\nu}$ and
$f_{\mu\nu}$. Clearly we have also to specify the definition of {\em
being in the same cluster}. This introduces a parameter that allow to
decide how deeply we want to test the clusterization features of the
data. We will introduce a threshold, that defines the refinement level
that we want to use to check our description.

\begin{figure}
\centerline{
\psfig{figure=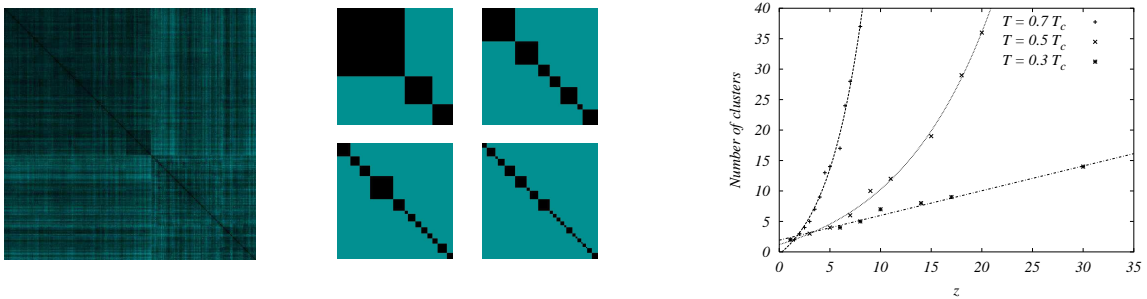,width=1.0\textwidth,angle=90}
}
\caption{On the left we plot the true distance matrix for a single
disorder sample at $T=0.5\ T_c$, and in the center four matrices
$f_{\mu,\nu}$ obtained for four different values of the threshold as
defined in equation (\protect\ref{E-HUBERT}). On the right we plot the
number of clusters $N_c$ versus $z$.  i.e. how the the number of
valleys depends upon the value of threshold we fix in order to test
the hypothesis. It turns out to be linear for small $T/T_c$,
exponential if $T\gtrsim T_c/2$ }
\label{F-HUBERT}
\end{figure}

We then have to define the a priori structure that we will compare to
the data.  Let us call $d_{\mbox{max}}$ the maximum distance (on the
hierarchical tree) among two configurations of our set: we say that
{\em two configurations belong to the same cluster if their distance
is smaller than a certain fraction of $d_{\mbox{max}}$, say than
${d_{\mbox{max}}}/{z}$}.  We show in the right part of figure
\ref{F-HUBERT} how the number of clusters $N_c$ depends on $z$. At
very low $T$ we find a linear dependence of $N_c$ over $z$, while at
values of the order of $\frac12 T_c$ $N_c$ grows faster than linearly.
In figure \ref{F-HUBERT} we also show, for one sample of the quenched
disorder, the true distance matrix $d_{\mu,\nu}$ and four different
matrices $f_{\mu,\nu}$ obtained with an increasing value of $z$ (from
the upper left corner going rightward and then to the lower line and
rightward again), $z=$ $4$, $8$, $12$ and $16$. The difference among
the structures that we are testing in the different cases is obvious.
The careful reader will be able to recognize by eye that the three
valley structure implied by the threshold level $z=4$ can indeed be
found in the raw distance data of the leftmost matrix.

The main ingredient needed for
analyzing the Hubert's
$\Gamma$ statistics is the correlation function
\begin{equation}
\Gamma=
\frac 1{M^2}\sum_{\mu=0}^M\sum_{\nu=0}^M
\frac{
\left(d_{\mu,\nu}-m_{D}\right)\left(f_{\mu,\nu}-m_{F}\right)
}{
{s_{D}\,s_{F}}
}\ ,
\label{E-GAMMA}
\end{equation}
where (for $X=D,F$, $x=d,f$)
\begin{displaymath}
m_X \equiv \frac 1{M^2}\sum_{\mu=0}^M\sum_{\nu=0}^M
x_{\mu,\nu}\quad\quad,\quad\quad
s^2_X \equiv \frac 1{M^2}\sum_{\mu=0}^M\sum_{\nu=0}^M
x^2_{\mu,\nu}-m_X^2\ .
\end{displaymath}
Let us say that when looking at the output of the clustering we
observe a value of $\Gamma$ equal to $\Gamma^*$. In order to estimate
if this value hints for the hierarchical structure being intrinsic to
the data we have used a number of tests. The first test amounts to
little more than checking if our procedures are correct: we take as
$H_0$ the randomness hypothesis, i.e. we compare our ordered distance
matrix to a matrix where the configurations are at random. We would
find that the configuration is not atypical only if our programming
was wrong. We compute an histogram $P(\Gamma|H_0)$, i.e. the
distribution of $\Gamma$ under the null hypothesis of randomness, by
evaluating
\begin{displaymath}
\Gamma(\pi)=
\frac 1{M^2}\sum_{\mu=0}^M\sum_{\nu=0}^M
\frac{
\left( d_{\mu,\nu}          - m_D \right)
\left( f_{\pi(\mu),\pi(\nu)}- m_F \right)
}{s_D\,s_F}\ ,
\end{displaymath} 
where the $\pi$ are random permutations of the $M$ configuration.  A
cluster is not consistent with the hypothesis $H_0$ (in this case the
hypothesis that configurations have not been ordered) if it is
``unusual''. In order to quantify this statement, we introduce an
indicator $\Delta$ defined as
\begin{displaymath}
  \Delta\equiv 
  \frac{\Gamma^*-\langle \Gamma\rangle}
   {\sqrt{\langle(\Delta\Gamma)^2}\rangle}
\end{displaymath}
where the value of $\Gamma$ that we have observed in our sample and
where the averages are taken with respect to the conditioned
probability distribution $P(\Gamma|H_0)$.  As expected we always find
a very high value of $\Delta$ for all reasonable values of the
threshold $z$ (i.e., say, values of $z$ that produce from two to order
hundred valleys): $\Delta$ is of order $10^{2}$ and that it is only
weakly dependent on the temperature (even at $T=\infty$ this test
tells that, yes, we had ordered the configurations, rejecting in this
way $H_0$ in a very clear cut way, since we are dealing with a large
matrix). As expected this procedure gives positive results both on the
original set of configurations and after applying the reversing
procedure described in section \ref{SS-REVERSE}.

The rest of the (more crucial) testing of the Hubert's $\Gamma$
statistics has been done on the set of reversed configurations, where
the $Z_2$ symmetry has been eliminated (see section \ref{SS-REVERSE}).
We will discuss it later on, after introducing same other important
objects and methods.

The second tool we use to establish whether the particular
hierarchical structure we find is the correct one is based on the
evaluation of the so called \emph{cophenetic correlation coefficient}
$\cal K$. It is defined as
$$
  {\cal K}\equiv
  \langle d\cdot d_C\rangle-\langle d\rangle \langle
  d_C\rangle\ ,
$$
where the cophenetic distance $d_C(\mu,\nu)$ is measured on the
dendogram (and because of that it is ultrametric by definition). For
example, in the case of Ward clustering, it is the quantity
defined in (\ref{E-WARD}). A high level of correlation of true
distance and cophenetic distance  implies that the data have an
intrinsic ultrametric organization. On the contrary a low level of
correlation suggests that a true ultrametric structure cannot be
detected. $\cal K$ is a natural measure of the ultrametricity build in
our data set.

If we try to analyze our original configuration set without removing
the $Z_2$ symmetry (each configuration $\cal C$ has a corresponding
configuration $\cal - C'$ which appears with the same probability)
we measure a high value of $\cal K$, always higher than $0.97$. 
Interpreting this result as a confirm of the detection of an
ultrametric structure would be  wrong: the $Z_2$ implies a very
primitive form of hierarchical organization (states are grouped in two
well separated sectors of the phase space) and on finite, medium size
volumes, this is what we are measuring.

\begin{figure}
\centerline{\psfig{figure=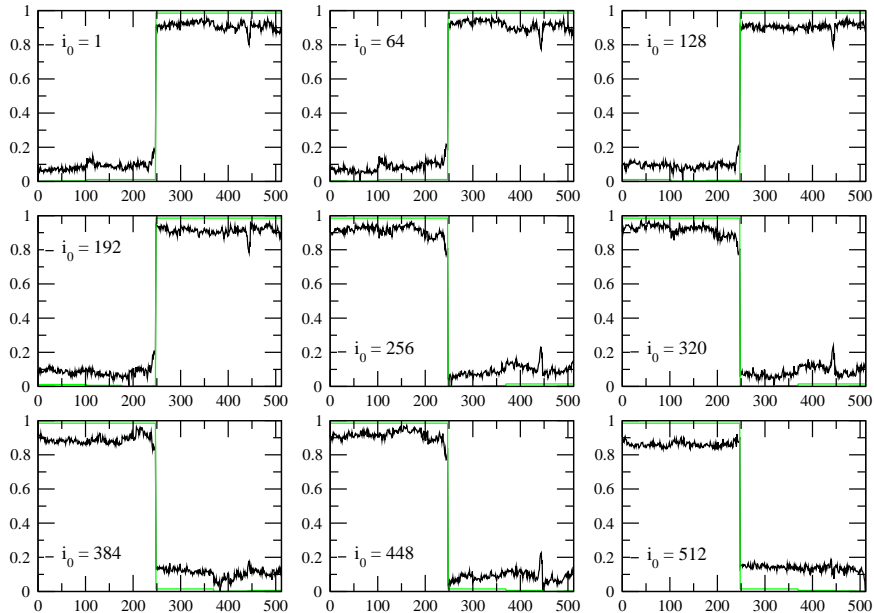,width=0.8\textwidth,angle=270}}
\caption{Plot of the true distance $d(i_0,j)$ (solid lines with wiggles) and
of the ultrametric cophenetic distance $d_C(i_0,j)$ (solid straight lines)
versus $j$ for different values of $i_0$.}
\label{cfr}
\end{figure}

One way to clarify this issue is to look at figure \ref{cfr}, where we
plot, for a given sample of the quenched disorder, at $N=512$ and low
temperature $T=0.3 T_c$, both the true distance $d(i_0,j)$ and the
cophenetic distance $d_C(i_0,j)$ as a function of $j$ for various
values of $i_0$. It is clear that the $Z_2$ symmetry makes the two
distances similar in a trivial way, by designing the same step: this
is the reason that makes ${\cal K}\lesssim 1$. The real physical
differences are in the wavy behavior of the true distance: it is its
difference from the constant behavior of the cophenetic distance that
has to be analyzed. This is what we will do in the next section.

We will now apply a spin reversal procedure that allows us to obtain a
set of configurations that have, in the infinite volume limit, a
positive definite mutual overlap. This is a very useful procedure
\cite{MAMAZU} that makes our set of configurations equivalent to a set
of configurations obtained in an infinitesimal magnetic field (without
the drawback of having to keep under control the smallness of the
field). Only after doing that we will come back to the evaluation of
the cophenetic coefficient $\cal K$. 

\subsection{The Reversing Procedure and Our Main Results\label{SS-REVERSE}}

\begin{figure}
\centerline{\psfig{figure=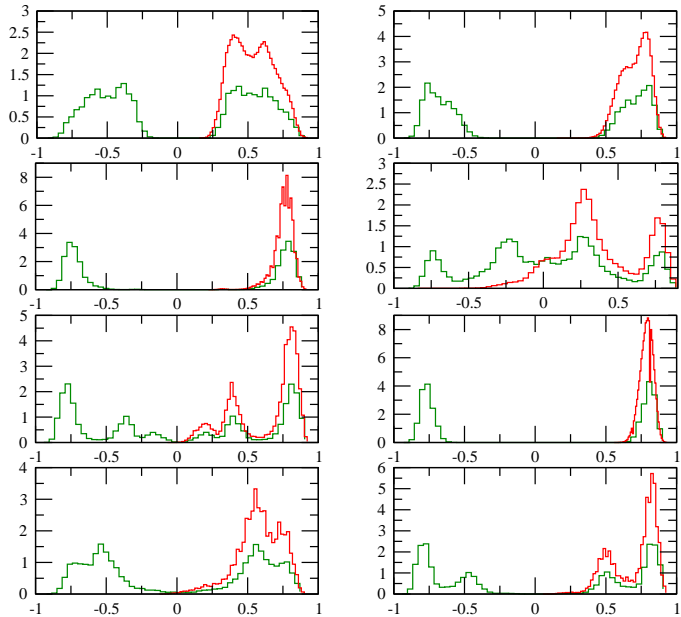,width=0.6\textwidth,angle=270}}
\caption{The probability distribution $P_J(q)$ 
for different realizations of the quenched disorder ($T=0.4$),
before and after applying the reversing procedure.
Here we use $M=512$ configurations of a $N=512$ spin system.}
\label{pdq}
\end{figure}

In the infinite volume limit the question of identifying in our set of
configurations two subsets, $|+\rangle$ and $|-\rangle$ is well
posed. After doing that we can flip all signs of the configurations in
$|-\rangle$, obtaining in this way a set of configurations with a
positive definite overlap. 

We use here the approach introduced in \cite{MAMAZU}. We take one
configuration as starting point, $\cal S$. We consider now a new
configuration, and if its overlap with $\cal S$ is negative we flip
it. For a third configuration we consider the average overlap with the
first two, and we flip it if this is negative. We do that for all
configurations. This procedure works quite well, and it can be
improved in a number of ways (for example we can repeat it by starting
from the new set and considering a different reference configuration
and a different order).

In figure \ref{pdq} we show the $P_J(q)$ for several samples, before
and after the reversing procedure. It is clear that the procedure
works quite well. The main problems are for samples where different
valleys are quite similar (we are on finite lattices and there are
intrinsic ambiguities that disappear in the thermodynamic limit). A
good example of a  troublesome samples is the second sample from the
top on the right, where the reconstructed $P_J(q)$ has, even after our
reversal procedure, a long tail at negative $q$ values. We have
verified (see also \cite{MAMAZU}) that when increasing the volume size
these spurious effects become smaller.

We have also found that a second effective approach to the separation
of the phase space is based on using the same clusterization procedure
we will eventually use for analyzing the hierarchical structure. We
first use clusterization (based for example on the Ward algorithm) to
identify the two $Z_2$ subsets. We then flip all spins of all
configurations of one of the two, and repeat the clusterization to
find a new (hopefully faithful) hierarchical structure. This second
approach gives results that are very similar to the ones of the first
approach \cite{MAMAZU} that we have discussed before: for example the
resulting $P_J(q)$ are basically indistinguishable.

In the following we will use spin configurations {\em ``reversed''}
using this technique.

\begin{figure}
\centerline{
\psfig{figure=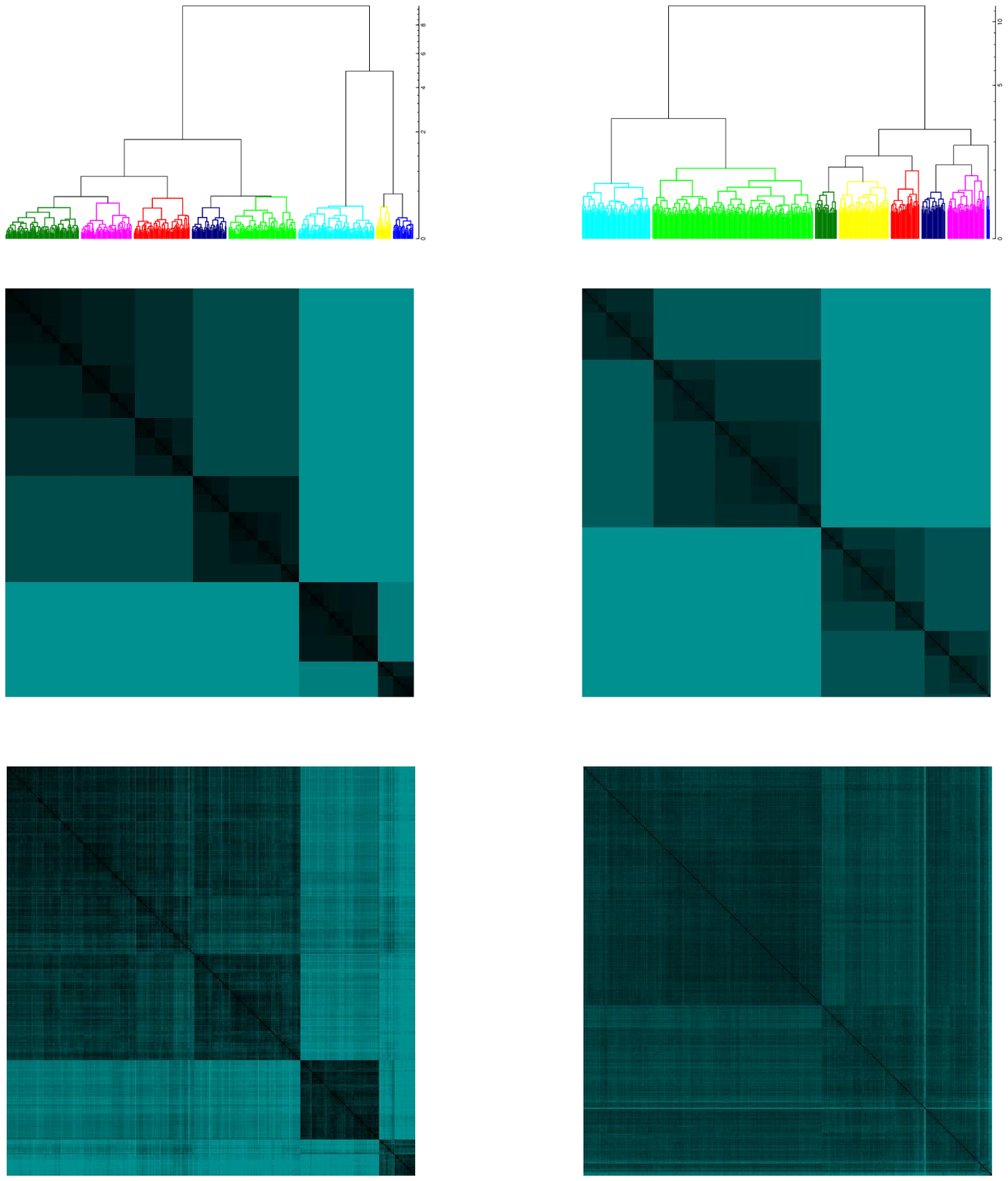,width=0.5\textwidth}
\psfig{figure=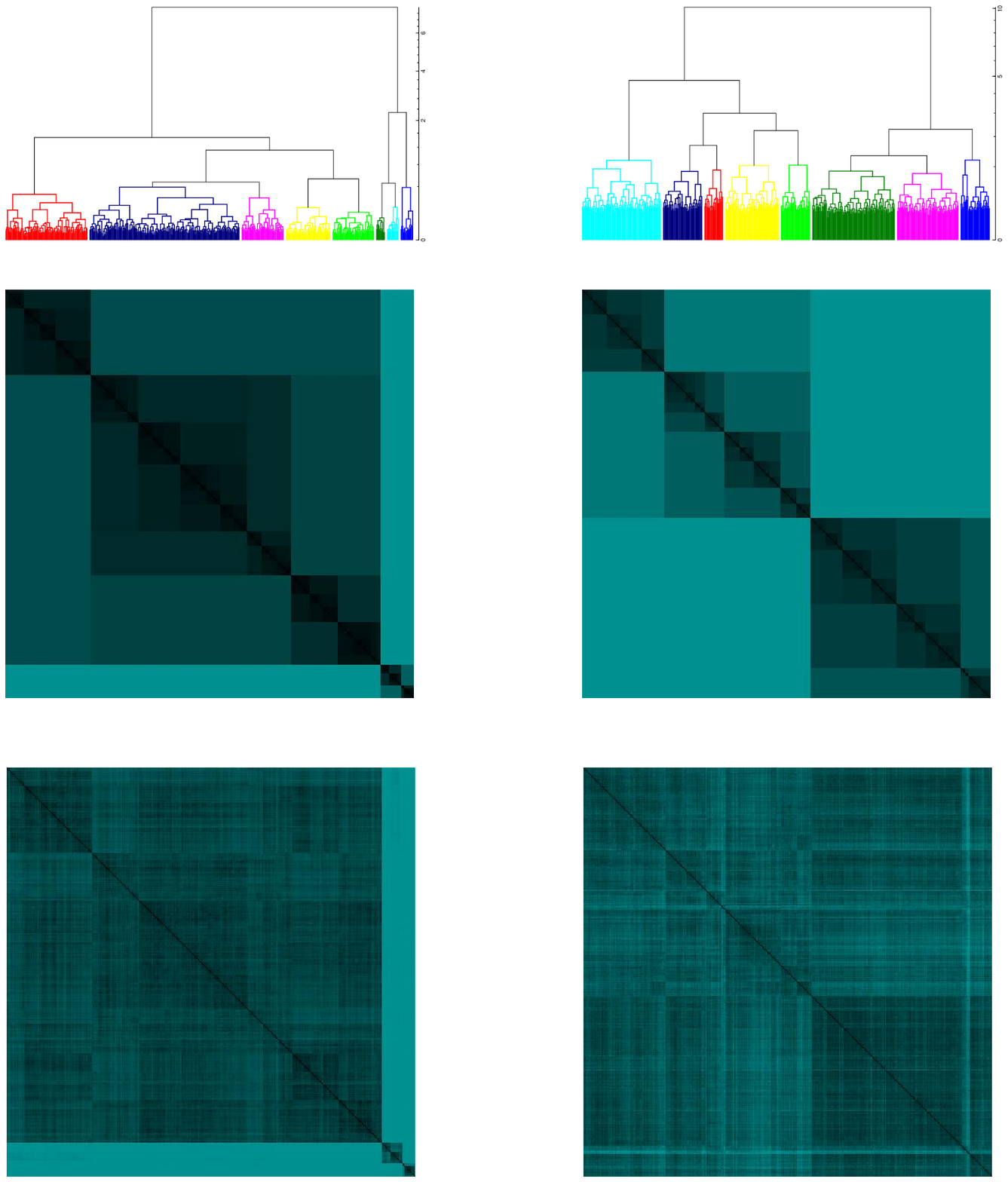,width=0.5\textwidth}
}
\caption{Proximity matrix for two $N=512$ samples in the left and
right parts of the plot (at $T=0.2\ T_c$ on the left for each of the
two samples and at $T=0.6\ T_c$ on the right for each of the two
samples) ordered according to the output of the clustering procedure
(i.e. as from the dendogram, in the bottom) and the corresponding
cophenetic matrix implied by the same dendogram (in the top).}
\label{A16}
\end{figure}

In figure \ref{A16} we show the proximity matrix for two $N=512$
samples (at $T=0.2\ T_c$ and at $T=0.6\ T_c$) ordered according to the
output of the clustering procedure (i.e. as from the dendogram) and
the corresponding cophenetic matrix implied by the same dendogram.

\begin{figure}
\centerline{\psfig{figure=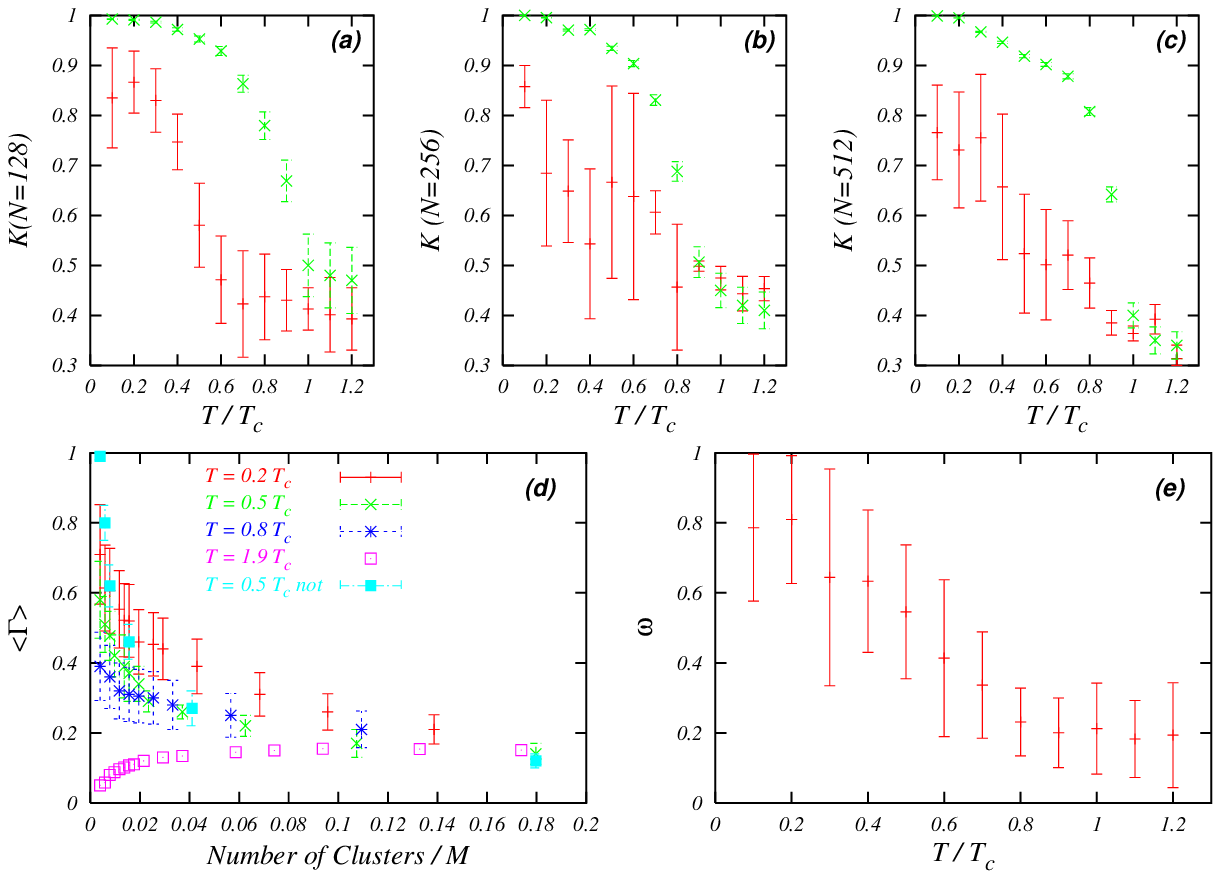,width=0.8\textwidth,angle=0}}
\caption{ In figures \protect\ref{corr}.a, \protect\ref{corr}.b and
\protect\ref{corr}.c we plot $\cal K$ as a function of $\frac{T}{T_c}$
for $N=128$, $N=256$ and $N=512$.  In figure \protect\ref{corr}.d we
plot $\langle\Gamma\rangle$ versus the assumed density of valleys,
i.e. the number of valleys divided times the number of configurations
$M$: a large difference from the high $T$ data implies a plausible
hypothesis.  In figure \protect\ref{corr}.e we compare single and
complete link clustering: see the text for further details.}
\label{corr}
\end{figure}

When the hierarchical, ultrametric structure is intrinsic to the data
set the matrices in the bottom line of figure \ref{A16} become equal
to the ones in the central line. Now that the accidental $Z_2$
symmetry has been removed we are able to look at the real, relevant
physical effects. We have investigated the issue in a systematic
way. We average over $20$ different quenched realizations of the
disorder, and analyze the system for different lattice volumes as a
function of the temperature. 

In figures \ref{corr}.a, \ref{corr}.b and \ref{corr}.c we plot $\cal
K$ as a function of $\frac{T}{T_c}$ for $N=128$, $N=256$ and $N=512$.
The upper sets of points with smaller errors are from the analysis
done {\em before} the spin reversal ($Z$), the lower sets of points
with larger error are from the analysis of the spin reversed
configurations ($R$). We have already discussed the fake detection of
ultrametricity induced by the $Z_2$ symmetry. We discuss now the data
obtained after removing the symmetry. In no cases a clear evidence for
the existence of a true ultrametric structure emerges. $\cal K$ is
always small, and for $T<T_c$ it does not even increase clearly with
$N$ (finite size effects are very large and uncontrolled). It is
interesting that in the set of $Z$ data the phase transition is
detected quite clearly (but, as we have explained, what we observed is
no connected to a hierarchical structure, but only to the
usual breaking of the $Z_2$ symmetry). At high $T$ values, for $T>T_c$
the $Z$ and the $R$ sets of data coincide: here there is one single
state.

This analysis shows clearly that on medium size lattices it is
impossible to detect more than hints toward a hierarchical structure:
in our mean field model we know that ultrametricity will eventually
emerge, but very large lattices are needed for that.

In figure \ref{corr}.d we try a further test to improve the level of
our quantitative understanding. We could phrase our goal by saying
that we are trying to understand how many valleys we can be sure are
present in the phase space (we repeat that since we are studying the
mean field Sherrington-Kirkpatrick theory in the Parisi broken phase
we know that asymptotically an infinite number of such valleys will
emerge). We go back to $\Gamma$ defined in equation \ref{E-GAMMA}.  At
different $T$ values we change the threshold value $z$ and monitor the
number of valleys we are building for a given $z$ value (this depends
on $T$: we have discussed this procedure when commenting figure
\ref{F-HUBERT}). We measure $\langle\Gamma\rangle$ and we plot it
versus the average number of valleys per sample (all data are for
reversed configurations, except for one set of non-reversed data at
$T=0.5 T_c$ that we plot for sake of comparison). We use the high $T$
($T=1.9 T_c$) curve as a reference curve, and we consider it as the
randomness threshold: if at a given temperature $T$ the value of
$\langle\Gamma\rangle$ is very different than the high $T$ value we
consider that as evidence for existence of this number of valleys. 

Using the hight $T$ limit as the reference line looks to us as a
sensible choice (we have already discussing that using unordered
matrix lines is basically just a check of the correctness of our
procedure). If, for example, we select a value of the $x$ variable
(number of clusters divided by $M$) $x=0.002$, that in the case of
$N=512$ assumes the presence of two valleys ({\em after} removal of
the $Z_2$ symmetry) we see that at low $T$ the data are quite
different from the high $T$ ones, suggesting that we are probably
already detecting this (quite low) level of organization. When we try
a threshold implying a larger number of valleys (already for example
for three of four valleys on our larger lattice, $N=512$) the data are
not far from the high $T$ ones, implying a failure in supporting the
hypothesis.

We will discuss figure \ref{corr}.e in the next section.

\subsection{Other Clustering Algorithms\label{SS-OTHER}}

As we have discussed in some detail in section \ref{S-CLUSTER} the
cluster reconstruction algorithm is defined by selecting the rule used
to join two elements at different levels of the partitioning, an to
update the distance matrix after each step of refining the
partitioning level.

In our analysis we have used the Ward scheme \cite{ward} (that updates
the distances as in equation \ref{E-WARD}): this is believed to be an
optimal choice when there is no information \emph{a priori} on the
data \cite{jaidub}.

Basic clustering algorithms are the {\em single link} scheme and the
\emph{complete link} one.  We will not enter here in many details (see
\cite{jaidub} for further information), but let us say that in the
single link scheme one just demands a weak connectivity to merge two
subsets, and joins them to form a new cluster as early as possible,
while in the complete link scheme the opposite happens, and subsets
are joined to form a new cluster ``as late as possible''. Both methods
have advantages and drawbacks. The crucial observation that we will
use now is that when a real hierarchical structure is present all
these methods end up to give the same result, and to reconstruct the
same classification.

In these two algorithms we have that, if as before $\rho$ and $\sigma$
merge to form the new cluster $\rho'$ for all other clusters $\tau$:
\begin{eqnarray*}
d_{\tau,\rho'}=&
\min\{d_{\tau,\rho},d_{\tau,\sigma}\}&
\;\;\;\;\;\textrm{(single link)}\ ,\\
d_{\tau,\rho'}=&
\max\{d_{\tau,\rho},d_{\tau,\sigma}\}&
\;\;\;\textrm{(complete link)}\ .
\end{eqnarray*}
The reason for the names is in the graph theory interpretation of the
algorithms \cite{jaidub}. As we have already said it is not difficult
to show that if the true distance matrix is actually ultrametric the
optimal permutation with respect to these two algorithms is be exactly
the same.

In this framework we have introduced a last test of the structure of
our data: we check how different are the output of the two algorithms
to try to understand if we can detect further hints for an emerging
ultrametric structure.  We have analyzed 20 samples at several
temperatures values, and we show in figure \ref{corr}.e the average
correlation between the two output distance matrices, that is
\begin{displaymath} 
\omega
\equiv \overline{\langle d_{SL}\cdot d_{CL}\rangle} \ .
\end{displaymath} 
The correlation is very high at low $T$, and decreases toward the high
$T$ value around $T\sim 0.8\ T_c$. Again, on medium large lattice sizes
we can detect hints toward an emerging ultrametric structure but we
cannot in any way get a clear cut answer.

\section{Clustering the Spins\label{S-SPINS}}

\begin{figure}
\centerline{\psfig{figure=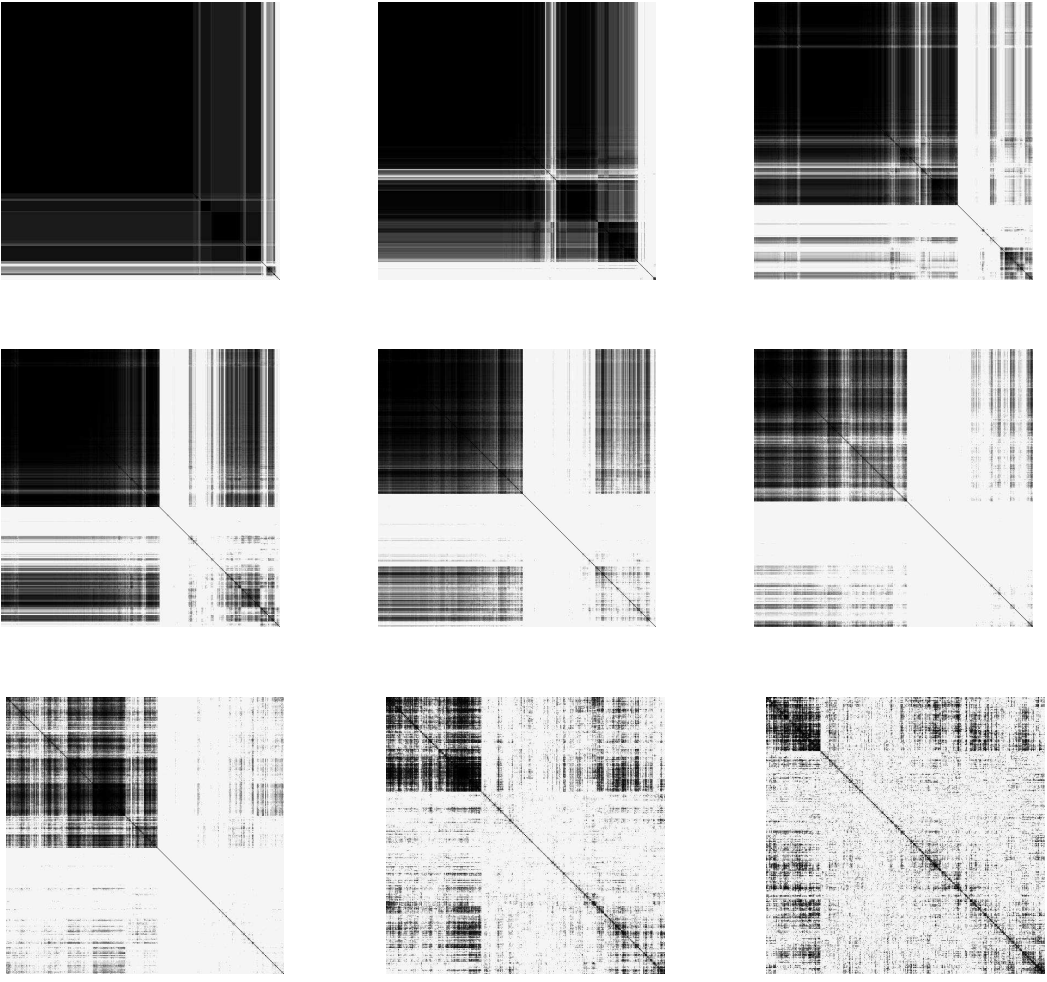,width=0.8\textwidth}}
\caption{Clustering the spins: for a given sample of the quenched
disordered couplings we look at the spins of our configurations as a
set done of $N=512$ elements (one per lattice site), each element
being a $M=512$ dimensional vector configurations (all the values
taken by the spin in the given site on our $M$ independent
configurations). After clustering these data vectors we plot the
distance matrix $d_{ij}$ between spin $i$ and spin $j$ according to
the ordering found in the cluster.  The plots correspond to $T=0.1\
T_c, T=0.2\ T_c, ...,0.9\ T_c$. At very low temperatures a large
($O(N)$) spin domain structure emerges. The structure disappears when
increasing the temperature.}
\label{spin}
\end{figure}

An interesting question (discussed in details in \cite{domany})
concerns a possible clustering of the {\em spins} of our system. 
The issue is clearly very relevant in the finite dimensional systems
studied in \cite{domany} where spatial structures can be very
relevant. Here, in mean field, there is no notion of distance, but
still spins can be aggregated in different groups that have different
degrees of correlation. 

We will look for the possible presence of some kind of structure (in
this case not hierarchical since there is no reason for this) now in
the space of the elementary spins instead than in configuration space.
In the analysis of configurations we were considering the $N\times M$
data matrix $\{\sigma_i^\mu\}$ as representing $M$ configurations,
where each data point was an $N$-dimensional vector.  Now we change
our point of view; we regard each of the $N$ spins as a data point,
that is as a vector in a $M$-dimensional space.  Since we expect
highly correlated spins to be in the same cluster, following
\cite{domany} 
we define
the distance between spin $i$ and spin $j$ as
$$
d_{ij}=1-c_{ij}^2\ ,
$$
where
$$
c_{ij}\equiv\langle \sigma_i\sigma_j\rangle 
\equiv\frac 1M\sum_{\mu=1}^M \sigma_i^\mu\sigma_j^\mu
$$
is the spin correlation matrix that we can evaluate using our 
spin configurations generated in a Monte Carlo run.

It is interesting to follow the evolution in temperature 
of the ordered spin matrix for a given sample: we show it in figure
\ref{spin}. At intermediate temperature values a large group of spin
is clearly very correlated: here $O(N)$ spins are grouped
together. This structure disappears at high $T$ values. It is
remarkable how this picture is similar to figure 11.d of the second
paper of reference \cite{domany}. This is a severe warning against
misleading interpretations of the data analysis: here we are in mean
field, and there are no spatial local domains.

\section{Conclusions\label{S-CONCLUSIONS}}

The configuration space of a $N$-spin system is a $2^N$-dimensional
space and it is very difficult to represent it in order to catch the
main physical features~\footnote{Only for limited purposes a principal
component analysis (PCA) can be adapted to help in this task
\protect\cite{domany}.}.  We have shown that cluster analysis allows
not only to visualize in a physically meaningful way the structure of
the configuration space, but also allows for quantitative testing of a
priori hypothesis about the structure of the data set.

We have discussed the role of the $Z_2$ symmetry of the system, and
how its removal is necessary to study the relevant physical
phenomena. Our main issue is that quantitative testing is mandatory to
make of clustering techniques an useful tool. We have introduced some
of these techniques by designing tests such to be useful in our
context of a (disordered) statistical mechanics context.

As a crucial benchmark we have analyzed the mean field theory in the
low $T$ replica broken phase, where we know that eventually, in the
infinite volume limit, a hierarchical structure of states emerges. We
are able to observe many hints toward the emerging of such structure,
but on the lattice sizes where we are able to work these indications
cannot be considered as unambiguous. Detecting ultrametricity is very
difficult, and demands very large lattice sizes: this turns out to be
true in mean field, and we expect it to be probably true also in
finite dimensional models, where the existence itself of mean field
like states is all to be checked. We believe that the findings and the
techniques that we have reported here will be important to use in the
finite dimensional context. As many other features (we have in mind
for example temperature chaos \cite{CHAOS}, that is very difficult to
detect numerically and emerges only at very high orders in
perturbation theory) ultrametricity emerges, already in mean field,
only on very large lattices.

We also believe it is important that in this ``quantitative'' approach
to clustering we have been able to introduce a natural way to consider
not only sample dependent but also disorder average quantities.

A next step is to apply, by continuing the work of \cite{domany},
these techniques to finite dimensional disordered systems (defined on
very large lattices!) on the one side and to glassy systems on the
other side: since here a crucial goal is to try to understand the
details of the spatial, time dependent organization of the system,
techniques like the ones introduced here could turn out to be very
useful.

\section*{Acknowledgments\label{S-ACK}}

We acknowledge the precious contribution of Loredana Correale to a
first phase of this work. We thank Eytan Domany and Peter Young for many
useful conversations that have motivated us toward this problem.


\end{document}